# New Revelation of Lightning Ball Observation and Proposal for a Nuclear Fusion Reactor Experiment

Domokos TAR

*M. of physics ETH-Zürich; Eichtlenstr.16, CH-8712 Stäfa*

**Abstract**
In this paper, the author brings further details regarding his Lightning Ball observation which were not mentioned in the first one. Additionally, he goes more into detail as to the three forces which are necessary to allow the residual crescent from the hydrodynamic vortex ring to shrink into a sphere.

Further topics are the similarities and analogies between the Lightning Ball formation's theory and the presently undertaken Tokamak-Stellarator-Spheromak fusion reactor experiments. A new theory and its experimental realisation are proposed as to make the shrinking of the hot plasma of reactors into a ball possible by the help of the so called long range electromagnetic forces. This way, the fusion ignition temperature could possibly be attained.

## 1. An eyewitness report

The author describes his observation of Lightning Ball in more details as already published in [1].

"In about 2 sec after the disparition of the lightning flash suddenly a very strong air-turbulence, a turning ring-cylinder appeared about 1.2 m over the ground FIGURE 1. I was very astonished: How it is possible that from apparently nothing a turbulent ring-cylinder appeared, like a miracle? I could see through that cylinder but nothing was to see at the end. It was like a tube of about 50 cm diameter outside and with an inner diameter of about 30 cm. The ring was formed by a lot of coaxially turning rings of moisty, dirty leaves. The axis of the ring was horizontal to the ground and it turned counter clockwise. In the middle of the ring it was nothing to see. The inner board had a sharp contour. The whole phenomenon lasted about 1-2 sec. After that time the ring suddenly disappeared and at the same moment an illuminated sphere appeared exactly in the middle of the ring. From the ring was nothing more to see and the sphere had a reddish tail. I could see the tail during about 0.1-0.4 sec. After that the ball appeared in full beauty and wandered with constant speed to the left in spite of the strong whirlwinds. My first thought was: *What strange phenomena exist in Nature!* "

In the second paper [2] was postulated that the cutting and shrinking of the hydrodynamic vortex ring into a crescent form is necessary for the formation of the ball FIGURE 2. At he ends of the crescent there are two equal opposed forces which press the crescent into a ball. There are two symmetry axes, the first one is perpendicular to the torus plane, the second one is the torus axis. But a sphere has 3 symmetry axes, which are all perpendicular to each other and cut themselves in the centre of the ball. The sphere has a higher symmetry. Nature endeavours to higher stability. The author believes that the inner volume of the cylinder-ring could have been slightly evacuated and that

135

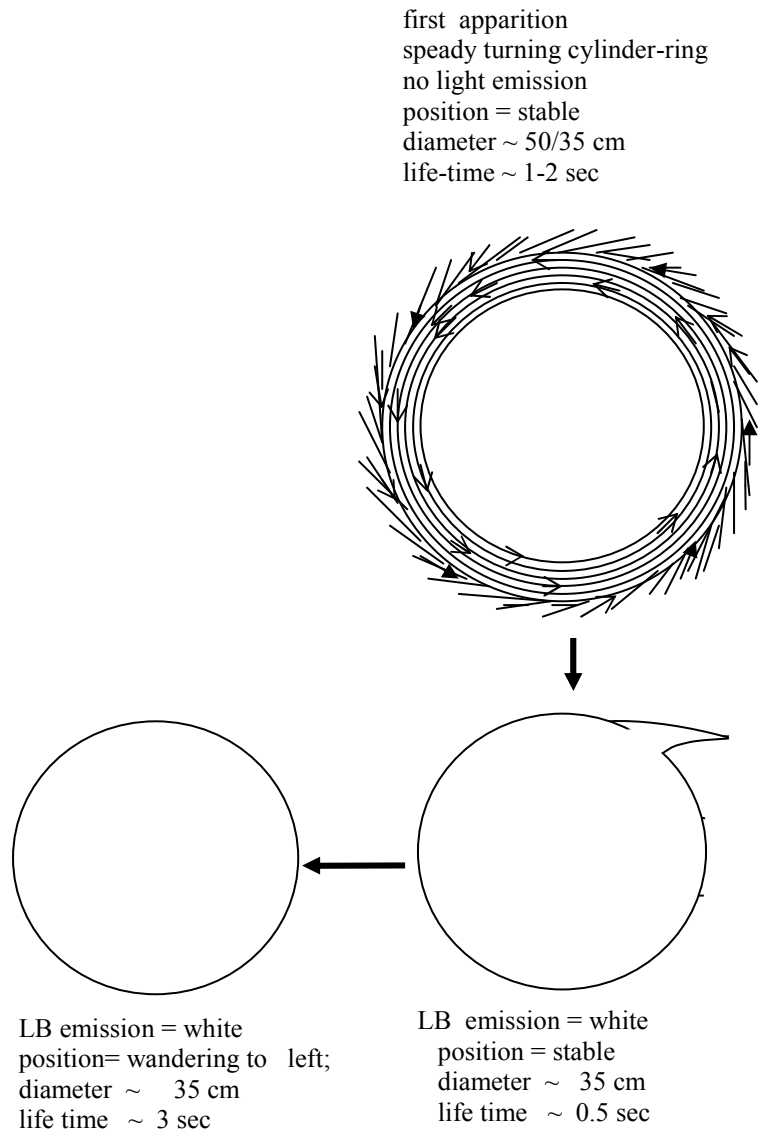

**FIGURE 1.** Creation of the Lightning Ball from a turning air-cylinder

the light emitted in the sphere was caused by the tribo-electroluminescence [2], [3] of the air. The transformation of the vortex ring into a turning cylinder should happen in a very short time (msec). Notice: a somewhat similar (but not the same) turning air cylinder had been fotographed in CH in a night [4].   The shrinking is illustrated in the model of FIGURE 3. The upper part of this figure shows the toroidal vortex ring in the cold state yet. The middle part is the crescent form at higher temperature after a successful cutting. The bottom part of the figure shows the final state of the crescent: a perfect sphere at high temperature.

Trying to explain his observation of the lightning ball formation, the author first thought to an analogy with the stopping of the lightning channel of ionized air-flow to the ground. Another phenomenon is the smoke vortex ring, produced by skilful smokers. In the elementary physics lectures smoke rings are demonstrated by striking a smoke box which has an opening. But in the lightning ball observation the spin of the vortex ring around its axis is just the opposite to a hydrodynamic air-flow and to the turning of a smoke ring. This is a proof that the LB's material does not come from the lightning channel but on another way. The author supposed [2] that the vortex created by the lightning was formed by a superposition of a shock (ultrasound wave over mach velocities) and by the normal sound wave at low velocity or formed by the explosion of the hot-compressed air, created by the return stroke just in the opposite direction.

136

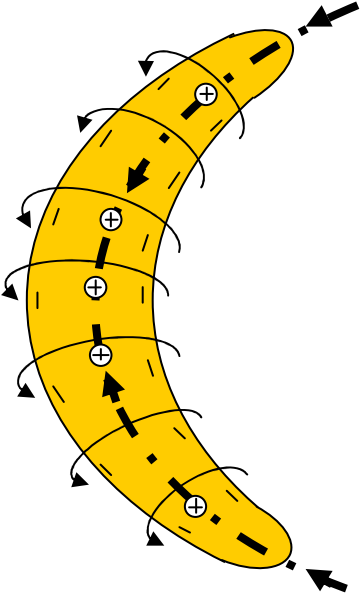

**FIGURE 2.** The plasma's crescent form is on the way of shrinking into a sphere by three forces

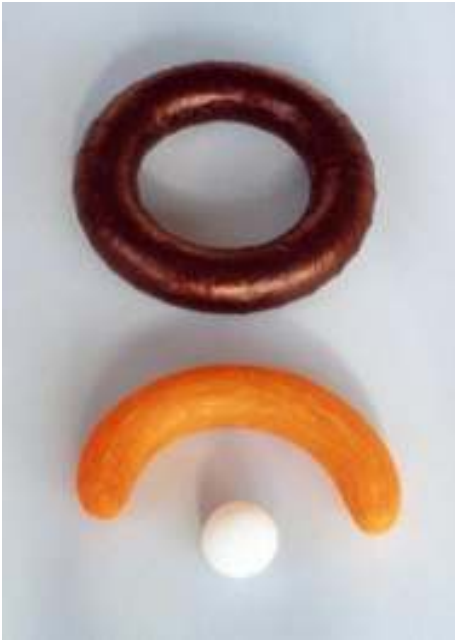

**FIGURE 3.** 3-D model of the Lightning Ball formation theory

D.Wells described his smoke ring observation in [5]. A smoke gun produced great smoke vortex rings. After that the ring was broken up, he observed an agglomeration of smoke into a blob with elliptical form. This observation served him to a completely geometrical approach to the formation of our solar

137

system with its planets and to arrive to a better approximation of the Titius –Bode law. This law gives the approximate radii of the planetary orbits in our solar system in relative units. Wells proposed that first of all different concentrically long cylinders were developed in the interstellar medium of plasma. These shortened later in a lot of rings all turning in the so-called accretions-sheet around the centre of galaxy. Later were these each broken up and agglomerated in different blobs, later to spheres, like the smoke concentration in his observation.

Now, to understand the difference between his geometrical model of the solar planets formation and the LB observation by the author we are considering the following facts:

**In the astronomy:**
1. there are gravitational forces.
2. these are central forces too between the sun and the planets.
3. the rings in the accretions-sheet have only toroidal rotation but no poloidal rotation.
4. at one point of each turning ring the plasma agglomerates into a planet.

**In the LB observation:**
1. there are long-range electromagnetic forces, but no gravitational forces.
2. there are no central forces in the vortex plane.
3. there are inertial moment forces.
4. the turning vortex has only a poloidal rotation, no toroidal rotation, it is like a smoke ring. but turning just in the opposite direction.

As a résumé we can see that the proposed LB theory is completely different from both theories: from the smoke vortex ring observation and Wells theory for the creation of the planets in our solar system. D. Wells had built a nuclear fusion reactor experiment called theta pinch [5]. At each end of the cylindrical setup was a "gun" generating two plasma vortex rings with opposed spins. They were fired each against the other with the aim that the two vortex rings would unify in only one vortex, due to the so called magnetic field reconnection. This is very uncertain in laboratory conditions. The experiment failed. R. Auchterlonie used the same principle in his patent application for a theta pinch, but in a system of double cylinder of toroidal configuration [6]. According to his theory, the cutting this double cylinder would result a vortex ring. That means that from a vortex cylinder-tube one would get a vortex ring again. This principle shows no forward step, because the result is again an unstable vortex. In practice this principle could not have been realized. However in the author's patent [14] there is a forward step, because the turbulent vortex ring should develop into a sphere, to a higher state of stability.

## 2. Photograph of a lightning Ball with two flashes

O. Prochnow published in 1928 a photograph of a LB with two flashes striking it (FIGURE 4 and [7]). The author believes in the genuinity of this photograph. This picture proves well the theory that LB and BL are two different phenomena [2]. If one examines this picture, the main question comes empor: Which was first the LB or the two flashes? The answer is simple: The ball was first, because the flashes are thicker at the board of LB, than they are at the beginning of the avalanches. This is very characteristic of each electrical avalanche discharges in Nature. It proves well the theory of the author that LB's with high energy get their energy from outside, from normal lightnings.

## 3.1. Comparison between the author's theory of LB formation and the Tokamak/ Stellarator/ Spheromak nuclear fusion reactor experiments; Proposal

If we draw up **the similarities** between the theory of the LB formation from a hydrodynamic vortex ring [2] and the different toroids of the nuclear fusion experiments (FIGURE 5 and [8]-[12]) we can conclude that:
1. Both are rings with cross section of circle or shaped circle form.
2. Both are turning around of two different rotation's axes.
3. Both have the property of expansion. At the fusion reactors the torus leads to a loop force which is in the direction to expand the plasma ring. This force is balanced by applying a vertical magnetic field which interacts with the toroidal current to give an inward force.
4. In both cases the vortex rings are geometrically not stable and they endeavour to a higher stable symmetry, the sphere.

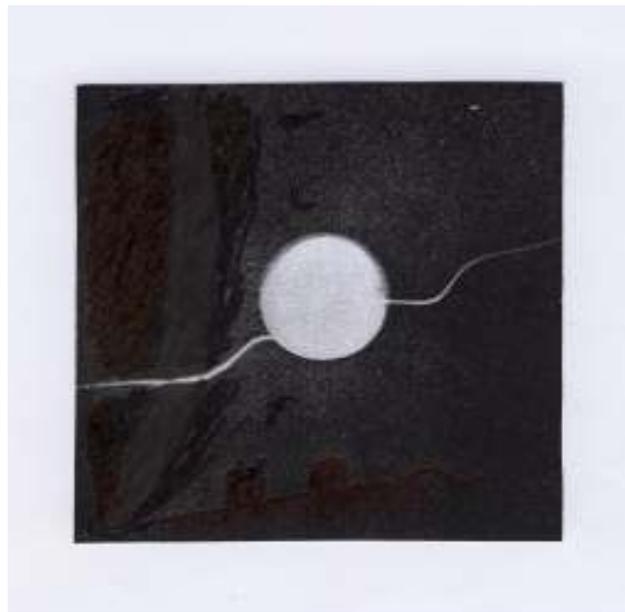

**FIGURE 4.** Photograph of two flashes striking into a Lightning Ball (photo Prochnow [7])

**The differences** are follows: the vortex ring of the LB theory is probably only superficial, because at the observation the ring was very hollow. Therefore the ball should be a hollow sphere, with a highly elastic surface. The surface is held together by the surface tension forces. However the whole volume of the fusion torus is more or less filled with the plasma.

**The EU patent application** [14], [13], [15] proposes that the torus of Tokamak/ Stellarator/ Spheromak fusion reactor experiments should be cut in its cross-section and so thanks to the so called long-range electromagnetic forces ([8] p.2) between the particles (collective behaviour), to allow the shrinking of the plasma into a sphere or at least into a smaller volume (FIGURE 3 and 5). But for the cutting (breaking) one should meet the best appropriate conditions to prepare the plasma, that the cutting succeeds. The cutting can be made by different proposed methods like: cutting by mechanical screens, diaphragms, magnetic mirror, disturbing the plasma with neutral ions injection, or by HF radiators which all have the purpose unifying the different plasma islands ( [10] p.372) into only one bunch, a sphere ( like a heavy fog becomes to a great water droplet).

### 3.2. Analysis of the magnetic field of the torus and plasma behaviour

The magnetic field is an infinite set of nested toroidal magnetic surfaces FIGURE 5. The field lines follow a helical path on their magnetic surfaces as they wind round the torus [8]-[12].The direction of the magnetic field changes from surface to surface. This shear of the magn. field has important implication for the stability of the plasma. Its twist is characterised by the safety factor q. The particles and ions fly in ant direction lines on helical paths, on which spirals are superimposed (Stellarator). After one tour on the circumference of the torus the particles do not fly on the same traces but shifted away. In sum the traces are very complicated. Therefore the whole movement mass flow is time variable and not stable. The plasma is not isotropic and not of the high density and temperature, what we would like.

139

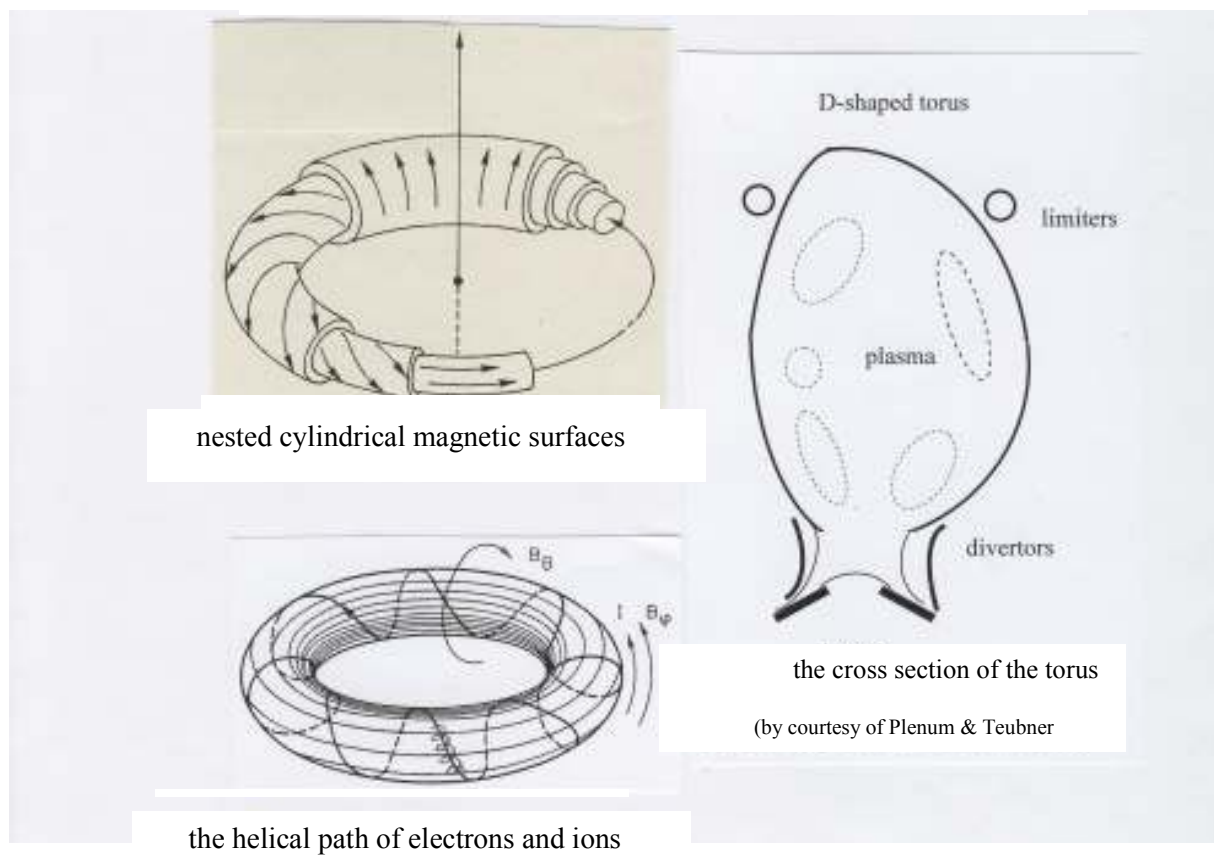

nested cylindrical magnetic surfaces

the helical path of electrons and ions

the cross section of the torus

(by courtesy of Plenum & Teubner)

**FIGURE 5.** The Tokamak magnetic confinement of plasma

All Tokamaks <u>with divertors and limiters</u> have the following property: At a well defined transition point the plasma temperature, density and energy confinement time suddenly improve by a factor of 2-until 6 (called H-mode state) and after that some parts of plasma break out and flow to the board resulting a cooling of the centre. This is a reversible process called the saw-teeths oscillations (Kadomtsev [10] p.342). Until now no explanation have been found for them. Are the saw-teeths oscillations not a signal that the plasma (on the way of getting hotter in a smaller volume) is for some reason disturbed? The H-mode (with higher temperature) of the saw-teeths appears only with mechanical divertors and limiters and therefore they are possibly very beneficial in this process? The falling back to the original state could possibly be prevented by placing screens outside of the axis. So the decrease of density could be stopped and allow the further rising of the central temperature.

We can consider the situation as follows. There is the two-fluid model of particles and the magnetohydrodynamic laws of the conservation of mass, momentum and energy are valid. Only the long-range electromagnetic forces act between the charges, without the short-range coulomb collisions. The physical picture: We imagine children playing hand in hand in a circle (wreath). They are turning around and in spite of the expanding forces they still are keeping their hands without loosing them. There are two different games: children are running around or they are staying at the same place, but still are tying to expand the diameter of the circle. Now, to the fusions reactor experiment. If the toroidal velocities of all particles (charges) are zero (no current flowing) and the plasma accidentally breaks in a poloidal plane, a bunch of nuclei automatically assembles at the opposite part of the torus. So the probability of the fusion collisions highly increases. The idea can be applicable to all kind of fusion reactor experiments, like: tokamak, stellarator, spherical tokamak, spheromak, levitated dipole experiment etc.

If we take the observation in Nature as an example and we compare it with a hydrodynamic flow, we can carry out the following measures for a good preparation of the plasma before breaking: The velocities of electrons and ions should be as high as possible in the poloidal direction and at the same time these velocities should be zero or as small as possible in the toroidal direction. This is proposed from a hydrodynamic consideration and comparison. In consequence of this statement : in supplement of the centripetal force in the cross section (poloidal plane) of the torus, two opposite toroidal forces are necessary at the end of crescent form that this can be shrink into a sphere Fig.2. This follows from the long-range collective behaviour. The velocities of electrons and ions can be adjusted, vary experimentally in both poloidal and toroidal direction. The resulted sphere has much more stability at the least surface by the greatest volume.

The proposed method of the mechanical cutting of a tokamak/stellarator torus is in principle the same as one would bend a linear magnetic mirror machine into a torus and replace the two magnetic coils at both ends by only one mechanical switch (screen) and rejoining both ends. It is well known, that the reflexions–efficiency of the mirror is very low, whereas this is by the mechanical switch nearly 100%.

### 3.3. The problems of the cutting

Placing mechanical screens into the plasma can cause a radical disruption of the plasma current and a rapid cooling. To counteract this effect we can propose the following measures: First of all a cutting must not necessarily be fully, but only partial. The pitch of helical field lines are tighter in the centreline of the plasma. Therefore the cutting screen can be like an Iris diaphragm (f-stop) in a photo-camera, which cuts only the board of the torus, leaving the central part free [13]-[15]. An other form could be like a jumping -umbrella which shadows only the central part of the torus and so lives the residual part free. Other forms of screens are also possible. The plasma is not an isotropic continuum, but is distributed in different plasma islands. Therefore an appropriate cutting or disturbing in a good way could unify many of these islands. A second aspect: It is possible that a very rapid cutting is not so harmful for the plasma. Anyway the tokamaks work in a pulsed operation. A breaking is possibly less harmful in a Stellarator or Spheromak. There is a third aspect. It is known, that a perfectly conducting first wall avoids the plasma from striking it and this causes an intern stability of plasma [11] pp.259, 302 , 334. Therefore it is possible, that a screen of well conducting material too, could make a cutting (separation) better.